\documentstyle[11pt]{article}

\newcommand{\nc}{\newcommand}
\nc{\be}{\begin{equation}}
\nc{\ee}{\end{equation}}
\nc{\bea}{\begin{eqnarray}}
\nc{\eea}{\end{eqnarray}}
\nc{\mc}{\multicolumn}

\begin{document}

\begin{titlepage}
\begin{flushright}
HUB-EP-97/29 \\
MS-TPI-97-5\\
\end{flushright}
\vspace{3ex}
\begin{center}

{\LARGE \bf
\centerline{$A_{+}/A_{-}$, $\alpha$,
$\nu$, and $f_{\rm s} \, \xi^3$ from}
\centerline{3D Ising Energy and Specific Heat}
}

\vspace{1.2 cm}
{\large M. Hasenbusch,}

\vspace{0.3 cm}
{\it Fachbereich Physik, Humboldt Universit\"at zu Berlin,} \\
{\it Invalidenstr.\ 110, 10099 Berlin, Germany} \\[2mm]
email: hasenbus@birke.physik.hu-berlin.de

\vspace{0.8 cm}
{\large K. Pinn,}

\vspace{0.3 cm}
{\it  Institut f\"ur Theoretische Physik I, Universit\"at M\"unster,} \\
{\it  Wilhelm-Klemm-Str.\ 9, D-48149 M\"unster, Germany} \\[2mm]
email: pinn@uni-muenster.de

\vspace{1. cm}
\end{center}
\setcounter{page}{0}
\thispagestyle{empty}
\begin{abstract} \normalsize

We analyse Monte Carlo data  for the energy and specific heat at and
close to the critical point of the 3D cubic Ising model.
{}From the finite size scaling of the energy $E$ and the specific heat $C$
at criticality we obtain the estimate $\nu = 0.6308(10)$. Furthermore,
one obtains precise estimates for the ``backgrounds'' (nonsingular parts) 
$E_{\rm ns}$ and $C_{\rm ns}$. 
Fitting solely off critical energy estimates to a scaling law, 
we find, 
depending on the choice of the reduced temperature, either 
$A_{+}/A_{-} = 0.550(12)$ and $\alpha=0.1115(37)$, or 
$A_{+}/A_{-} = 0.567(16)$ and $\alpha=0.1047(48)$.
Including information from the data at $T_c$, we obtain 
the estimate $A_{+}/A_{-} = 0.560(10)$. 
We also determine the universal combination
$f_{\rm s} \, \xi^3$ in both phases.

\end{abstract}
\nopagebreak
\vspace{1ex}
\end{titlepage}
\newpage
The universal amplitude ratio $A_{+}/A_{-}$
of the 3D Ising universality class 
(for a precise definition see eq.~(\ref{Cansatz}) below)
still seems subject to some uncertainty. For a general
discussion of  the difficulties one encounters when trying to estimate
$A_{+}/A_{-}$ from high and low temperature expansions see
ref.~\cite{LiuFisher}.  A compilation of some results in the literature
will  be given in the last section of this article. For a general
introduction to universal critical-point amplitudes see
e.g.~\cite{privman_amp}.

We here present a calculation  of $A_{+}/A_{-}$ based on Monte Carlo
data for the energy of the 3D Ising model.  Furthermore, we obtain 
fairly precise estimates of other  quantities, like the exponents $\nu$
and $\alpha$, nonsingular parts of energy and specific heat, and of the
universal combination $f_{\rm s} \xi^3$ on both sides of the transition.

Consider the 3D Ising model on the simple cubic lattice of size $L
\times L \times L$, with  periodic boundary conditions.  The Hamiltonian
is 
\be\label{isiham}
H = - \sum_{<x,y>} s_x s_y  \, , \quad
s_x = \pm 1 \, .
\ee
The sum in eq.~(\ref{isiham})
is over all (unordered) nearest neighbour pairs of sites
in the lattice. The partition function is
\be
Z = \sum_{\{s\}} \, \exp \left( - \beta H \right) \, .
\ee
Here, the summation is over all possible configurations of the Ising
spins. The pair interaction is normalised such that $\beta=1/(k_B T)$,
where $k_B$ denotes Boltzmann's constant, and $T$ is the temperature.

At a critical coupling $\beta_c=0.221 6544(6)$~\cite{talapov} the  model
undergoes a second order phase transition. For $\beta > \beta_c$, the
system shows  spontaneous breaking of reflection symmetry.

The free energy density (free energy per link) is defined by
\be
\label{freeenergy}
f  = - \frac{1}{3 L^3} \ln Z \, .
\ee
We define the energy (per link) as  the derivative of $f$ with 
respect to $\beta$, 
\be
\label{energy}
E = -  \frac{d}{d\beta} f = 
    - \frac{1}{3 L^3} \langle  H  \rangle \, . 
\ee 
The specific heat is defined as the derivative of $E$ with respect 
to $\beta$, 
\be
\label{heat}
C = \frac{d}{d\beta} \, E = 
\frac{1}{3 L^3} 
\left( \langle H^2 \rangle - \langle H \rangle^2 \right) \, . 
\ee
Note that choosing other definitions, like putting a minus sign
in eq.~(\ref{energy}) or substituting a $d/dT$ instead
of the $d/d\beta$ in eq.~(\ref{heat}) leads to trivial
factors and/or signs in the definitions and results to be stated below.

The specific heat is singular at the critical point. Close to 
$\beta_c$ it is expected to behave like 
\be
\label{Cansatz}
C \simeq C_{\rm ns} + C_{\rm s} \, , 
\ee 
where $C_{\rm ns}$ is an analytic function of $\beta$ at $\beta_c$.
The singular part is 
\be 
C_{\rm s} \simeq A_{\pm} \, |\, t \, |^{-\alpha}  \, , 
\ee 
where 
\be 
t = 1 - \frac{\beta}{\beta_c} \,  
\ee 
is the reduced temperature. $A_{+}$ and $A_{-}$ denote the amplitudes of
the singular  part in the symmetric ($t>0$) and broken ($t<0$) phase,
respectively. $\alpha$ is the specific heat exponent.
The singularity of the specific heat implies a non-analytic behaviour of
the energy $E$ and the free energy density $f$.  Some details will
be given in the next section.

\section{Scaling and Finite Size Scaling}

For a general introduction to finite size scaling theory, see, e.g.,
ref.~\cite{privman_fss}. In order to discuss the non-analytic behaviour
of the free energy density it is useful to split it into
an analytic (nonsingular) and a singular part, 
\be 
f = f_{\rm ns} + f_{\rm s} \, . 
\ee
Renormalization group arguments lead to the following finite size
scaling ansatz for the singular part of the free energy 
density for lattices with periodic boundary conditions~\cite{cardy}, 
\be
\label{wichtig}
 f_{\rm s} \, \xi^d \simeq g(\xi/L) \;\; ,
\ee
where $f_{\rm s}$ is taken in the finite volume, while $\xi$ is the
correlation length defined in the thermodynamic limit. $g(\xi/L)$ is a
universal function. In the following we discuss the two extremal cases
of the thermodynamic limit and the finite size scaling exactly at the
critical point.  

The thermodynamic limit is characterised by $\xi/L = 0$.
Inserting the scaling ansatz $\xi \sim t^{-\nu}$ into eq.~(\ref{wichtig})
for $\xi/L=0$ we obtain 
\be
 f_{\rm s} \sim t^{d\; \nu} \, . 
\ee
By differentiation with respect to $\beta$ we arrive at 
\be
 E_s \sim t^{d\; \nu-1}
\ee
and
\be
 C_{\rm s} \sim t^{d\;\nu-2} \;\; .
\ee
The last equation implies the so called hyperscaling relation
$\alpha = 2-d \; \nu$.

In order to discuss finite size scaling at the critical point it is
useful to reparametrize eq.~(\ref{wichtig}) as 
\be
 f_{\rm s} L^d \simeq h(L/\xi) \, , 
\ee
with $h(L/\xi) = (L/\xi)^d \;\; g(\xi/L)$.
Inserting the scaling law $\xi \sim t^{-\nu}$
we obtain
\be
 f_{\rm s} \simeq L^{-d} \; \tilde h(L^{1/\nu} t) \, , 
\ee
and, by differentiation with respect to $\beta$, 
\be
 E_s \sim L^{-d+1/\nu} \; \tilde h'(L^{1/\nu} t) 
\ee
and
\be
 C_{\rm s} \sim L^{-d+2/\nu} \;  \tilde h''(L^{1/\nu} t)  \;\; .
\ee
For the critical temperature $t=0$ this means
\be
\label{finiteE}
 E_s \sim L^{-d +1/\nu} \, , 
\ee
and
\be
\label{finiteC}
 C_{\rm s} \sim L^{-d +2/\nu} \;\; .
\ee
In our numerical study we approximated the nonsingular part of
the free energy density by its Taylor expansion, 
truncated at second order,
\be
f_{\rm ns} \simeq  F_{\rm ns} - E_{\rm ns} \; (\beta-\beta_c) 
- \frac12 \, C_{\rm ns} \;\; (\beta-\beta_c)^2  \;\; ,
\ee
where $F_{\rm ns}$, $ E_{\rm ns}$ and $C_{\rm ns}$ are the nonsingular
parts of  the free energy density, the energy density and the specific
heat at  the critical point, respectively.

\section{MC Simulations}

\subsection{Simulations at $\beta_c$}

We simulated the model at  $\beta_c=0.2216544$ on lattices of size
$L=12$ up to $L=112$. For the simulation we employed the single cluster
algorithm. The updating between two measurements consisted of  a number
of clusters, ranging between 5 and 50, and  a single Metropolis sweep. 
The total number of measurements was several millions for the smaller
lattice and some hundred thousands for the larger systems. 

We measured the energy $E$, the specific heat $C$ and the 
derivative of $C$ with respect to the inverse temperature $\beta$.
Our results for $E$ and $C$ are summarised in table~\ref{attc}.

\begin{table}
\begin{center}
\begin{tabular}{|r|l|l|}
\hline
\mc{1}{|c}{$L$} &
\mc{1}{|c}{$E$} &
\mc{1}{|c|}{$C$}    \\
\hline
 12 & 0.352212(10) & 11.0572(24) \\ 
 16 & 0.344859(16) & 12.2103(58) \\ 
 20 & 0.340931(7)  & 13.1588(47) \\ 
 24 & 0.338489(12) & 13.9138(95) \\ 
 28 & 0.336873(6)  & 14.5920(55) \\
 32 & 0.335721(6)  & 15.1921(74) \\ 
 36 & 0.334882(6)  & 15.7199(84) \\  
 40 & 0.334233(7)  & 16.222(19) \\ 
 44 & 0.333735(6)  & 16.652(11) \\ 
 48 & 0.333302(10) & 17.075(25) \\    
 56 & 0.332701(8)  & 17.800(21) \\
 64 & 0.332286(9)  & 18.483(30) \\
 72 & 0.331954(9)  & 19.059(41) \\
 80 & 0.331720(7)  & 19.617(32) \\
 96 & 0.331365(8)  & 20.517(65) \\
112 & 0.331145(8)  & 21.439(80) \\
\hline
\end{tabular}
\parbox[t]{.85\textwidth}
 {
 \caption[attc]
 {\label{attc}
Results for the energy $E$ and the the specific heat $C$ at  
$\beta_c=0.2216544$ for various lattice sizes $L$.
 }
 }
\end{center}
\end{table}

We fitted our data for the energy and the 
specific heat according to the ans\"atze
\be
E=E_{\rm ns} + {\rm const}_E \; L^{-d +1/\nu}
\ee
and
\be
C=C_{\rm ns} + {\rm const}_C \; L^{-d +2/\nu}
\ee
that are motivated by eqs.~(\ref{finiteE}) and~(\ref{finiteC}),
respectively.
The results are summarised in table~\ref{atfit}.
The $\chi^2$ per degree of freedom becomes smaller
than one if only lattices with $L\ge 20$ are included in the fit.

\begin{table}
\begin{center}
\begin{tabular}{|l|c|l|l|l|l|l|l|}
\hline
data & 
$L_{min}$ & $X$ & $\nu$ &  ${\rm const}_E$  & $E_{\rm ns}$  &
${\rm const}_C$ & $C_{\rm ns}$ \\
\hline
\hline
energy & 12      &  1.04  &0.6280(5)&0.7276(25)& 0.330190(7)&  & \\ 
data   & 16      &  1.13  &0.6282(10)&0.729(5) & 0.330192(9)&  & \\
only   &20      &  0.86  &0.6296(12)&0.737(7) & 0.330200(9)&  & \\
\hline
\hline
specific    &12      & 1.42 & 0.6380(7)& & & 20.9(8)  & -18.2(9)  \\ 
heat data   &16      & 1.35 & 0.6365(13)& & & 19.3(1.2)&-16.4(1.4) \\
only        &20      & 0.45 & 0.6329(16)& & & 16.1(1.3)&-12.9(1.4) \\
\hline
\hline
both & 12 & 5.6 &0.6316(4)&0.7440(20)&0.330229(5)&15.32(30)&-12.13(33) \\
data & 16 & 2.25 &0.6315(8)&0.7460(40)&0.330218(7)&15.15(54)&-11.83(61) \\
combined&20 &0.75&0.6308(10)&0.7431(52)&0.330209(8)&14.58(66)&-11.12(76)\\
\hline
\end{tabular}
\parbox[t]{.85\textwidth}
{
\caption[atfit]
{\label{atfit}
Results of fits of the energy $E$ and the the specific heat $C$ at 
$\beta_c=0.2216544$. The upper part gives results from energy data
only, the following three rows state results from specific heat data only,
while the lower part refers to fits where both sets of 
data were combined. Only data of simulations with lattice size greater or
equal to  $L_{min}$ were used for the fits. $X$ denotes $\chi^2$ per 
degree of freedom.
}
}
\end{center}
\end{table}
One observes that the result for $\nu$ obtained from  the energy is
smaller than that from fitting the specific heat.   However, when
discarding data from small $L$ the estimate of $\nu$ from the energy
increases, while that from the specific heat decreases.

Next we fitted the data for the energy and the specific heat 
simultaneously. We checked that the cross-correlation of the two 
quantities is small compared to the geometric mean of the variances of
the two quantities. Therefore it is justified to treat for simplicity
the data as independent. When all data are included into the fit the
$\chi^2$ per degree of freedom  becomes unacceptably large.  Discarding
again the $L=12$ and $L=16$ data, the fits become  very good.  It is 
interesting to note that the value for $\nu$ changes only little when 
$L=12$ and $16$ are discarded from the fit. 

In order to check the dependence of our result on the value of the
critical coupling, we repeated the fit for $\beta=0.221 6538$ and 
$\beta=0.221 655$.  The values for the energy and the specific heat at
these $\beta$-values were obtained from first order Taylor expansion and
the numerically determined values of the derivatives. Taking into
account the error induced by the uncertainty of  $\beta_c$ we arrive at
the final estimates $E_{\rm ns}=0.330209(14)$, $C_{\rm ns} = -11.1(8)$
and $\nu=0.6308(10)$ obtained from the combined energy and specific heat
fit with $L\ge20$.  Here only statistical errors are given. It is
difficult to quantify systematic errors due to corrections to scaling. We
tried to use an ansatz that includes a leading correction to  scaling
term
\be
E_s \sim L^{-d +1/\nu} \;\; \left(1 + c \; L^{-\omega} \right) \, ,
\ee
with the correction to scaling exponent
$\omega=0.81(5)$~\cite{talapov,zinn_guillou}.
It turned out that the amplitude of the correction to scaling term 
was consistent with zero within errorbars. The estimate of $\nu$
was $\nu=0.631(4)$ when all data for the energy and the specific heat
were included in the fit.

Since the estimate of $\nu$ obtained from the the fit without corrections
to scaling is in nice agreement with recent results given in the 
literature (for an overview see~\cite{bloete}),
we also regard the estimates for $E_{\rm ns}$ and $C_{\rm ns}$, 
which will be used in the following, as reliable.

\subsection{Simulations at $\beta \ne \beta_c$}
\label{xxx}
Next we simulated the model at temperatures below and above the 
critical temperature, such that results for the thermodynamic limit 
could be obtained.
The resulting Monte Carlo estimates for $E$ are fitted to
\be
\label{Eansatz}
E \simeq E_{\rm ns} - C_{\rm ns} \, \beta_c \, t  
\, \mp \,  A_{\pm} \, \beta_c \, 
\frac{ |\, t\, |^{1-\alpha}}{1-\alpha}  \, , 
\ee 
which is obtained by integration of eq.~(\ref{Cansatz}).
For the simulations in the symmetric phase and for part of 
the simulations in the broken phase 
we used the single cluster algorithm~\cite{ulli} 
combined with a standard local Metropolis update. 
A typical mixture was 20 cluster updates plus 
a single Metropolis sweep, followed by a measurement of observables.
The total number of measurements was typically 
of order a few hundred thousands up to two millions. 
Part of the results for the broken phase were obtained 
in the course of an other project~\cite{hpisiface}, 
using a demon program coded in multispin fashion.
For details of these simulations we refer to~\cite{hpisiface}. 

\begin{table}
\begin{center}
\begin{tabular}{|l|l|rc|}
\hline
\mc{1}{|c}{$\beta$} &
\mc{1}{|c}{$E$}     &
\mc{2}{|c|}{$L$}    \\
\hline
0.2189088 & 0.311775(5)  &  96  & T \\
0.21931   & 0.313849(14) &  80  & T \\
0.2197088 & 0.315949(6)  &  96  & \\ 
0.2202    & 0.318742(11) & 100  & \\ 
0.2204    & 0.319958(6)  &  96  & \\
\hline 
0.2205    & 0.320587(9)  & 128  & \\
0.2205    & 0.320592(9)  &  96  & \\
\hline 
0.2206    & 0.321230(11) & 128  & \\
0.2206    & 0.321220(8)  &  96  & \\
\hline
0.2207    & 0.321887(9)  & 128  & \\
0.2207    & 0.321900(8)  &  96  & \\
\hline
0.2208    & 0.322581(9)  & 128  & \\
0.2208    & 0.322593(9)  &  96  & \\  
\hline
0.2209    & 0.323280(5)  &  128  &   \\
0.2209    & 0.323301(8)  &   96  & F \\  
\hline 
0.2210    & 0.323995(10) &  128  &   \\  
0.2210    & 0.324040(9)  &   96  & F \\ 
\hline 
\hline 
0.2220    & 0.340001(12)  & 128  & \\
0.2220    & 0.340001(34)  &  96  & \\
\hline 
0.2221    & 0.342368(19)  &  96  & \\ 
0.2222    & 0.344592(35)  &  96  & \\                                       
0.2224    & 0.348911(35)  &  96  & \\
0.2226    & 0.353134(36)  &  96  & \\
0.2228    & 0.357057(36)  &  96  & \\
0.2229    & 0.358935(47)  &  64  & \\
0.2230    & 0.360972(31)  &  96  & \\     
0.2234    & 0.368280(28)  &  96  & T \\
0.2236    & 0.371725(27)  &  96  & T \\
0.2238    & 0.375248(27)  &  96  & T \\
0.2240    & 0.378615(26)  &  96  & T \\
\hline
\end{tabular}
\parbox[t]{.85\textwidth}
 {
 \caption[Energy estimates for the 3D Ising model]
 {\label{eneres}
  Monte Carlo results for the energy of the 3D 
  off critical Ising model. 
  A ``T'' in the last column means that the corresponding data
  is not used for the fits because of its too large 
  reduced temperature. Exclusion of the fits because
  of finite size effects is indicated by an ``F''.
 }
 }
\end{center}
\end{table}
Our results for $E$ are displayed in table~\ref{eneres}.
The $\beta$-values were chosen in the range 0.218909 to 0.224.
The corresponding reduced temperature covers the interval from 
0.0124 to -0.011.  
The typical lattice sizes were 96 and 128. We convinced ourselves
that we always reached the thermodynamic limit within the 
numerical precision.
In the table we marked those values 
that were discarded because of finite size effects by an ``F''.
Data that were excluded because of a too large distance from
criticality are marked with a ``T'' (see below). 

We then made two types of fits: We first fixed $\beta_c$ and $\alpha$, 
and fitted $E_{\rm ns}$, $C_{\rm ns}$, $A_{+}$, and $A_{-}$.
Then we only fixed $\beta_c$ and fitted all the other 
parameters in eq.~(\ref{Eansatz}).
In both cases, we used in addition to the reduced temperature $t$ 
an alternative definition, 
\be
t'= \frac{\beta_c}{\beta}-1 \, . 
\ee
Comparing the fit results from the two definitions should give 
us an estimate of systematic effects that stem e.g.\ from the 
inclusion of data that have too large $t$ or the 
neglection of subleading terms in eq.~(\ref{Eansatz}).
The results for the fit parameters are summarised in table~\ref{fitres}.

\begin{table}
\begin{center}
\begin{tabular}{|l|c|r|l||l||l|l|l|}
\hline
\mc{1}{|c}{$\alpha$}       &
\mc{1}{|c}{type}           &
\mc{1}{|c}{$A_{+}$}        &
\mc{1}{|c||}{$A_{-}$}        &
\mc{1}{|c||}{$A_{+}/A_{-}$}  &
\mc{1}{|c}{$E_{\rm ns}$}   &
\mc{1}{|c|}{$C_{\rm ns}$}  \\
\hline
0.100 f &$t$&11.194(56)& 19.068(42)& 0.5871(16)& 0.33037(1)& -12.561(96)\\
   & $t'$ & 11.147(51)& 19.132(41)& 0.5826(15)& 0.33033(1)& -12.544(88)\\
\hline 
0.104 f&$t$& 10.374(52)& 18.076(39)& 0.5739(17)& 0.33034(1)& -11.549(91)\\
       &$t'$& 10.329(47)& 18.140(38)& 0.5694(15)& 0.22030(1)& -11.545(85)\\
\hline 
0.108 f&$t$&  9.625(49)& 17.160(37)& 0.5609(17)& 0.33031(1)& -10.613(89)\\
      &$t'$&  9.582(45)& 17.224(36)& 0.5563(15)& 0.33027(1)& -10.601(88)\\
\hline
0.112 f&$t$&  8.940(47)& 16.311(34)& 0.5481(18)& 0.33028(1)& -9.743(85)\\
       &$t'$&  8.900(42)&16.375(33)& 0.5435(15)& 0.33024(1)& -9.733(78)\\
\hline 
\hline 
0.1115(37)& $t$ & 9.03(63)&16.42(77)&0.550(12)&0.33029(3)& -9.86(80)\\
0.1047(48)&$t'$&10.19(96)&17.97(1.17)&0.567(16)&0.33030(4)& -11.37(1.17)\\ 
\hline 
\end{tabular}
\parbox[t]{.85\textwidth}
 {
 \caption[Fit results]
 {\label{fitres}
 Results for the fit parameters of eq.~(\ref{Eansatz}). An ``f'' in 
 the first column means that the corresponding parameter was 
 kept fixed to the quoted value. $t$ and $t'$ indicate the definition
 of reduced temperature that was employed in the fit.
 }
 }
\end{center}
\end{table}

We first started taking all the data of table~\ref{eneres}. However, it
turned out that in order to have fits with a  reasonable level of
confidence, we had to discard the  data marked with a ``T''. The fits
with the remaining data  (the results of which are quoted in
table~\ref{fitres})  had a $\chi^2$ per degree of freedom of 0.9 to 1.2.

There is a systematic difference  of the fits with the two different
definitions of the reduced  temperature. In case of the fits with fixed
$\alpha$ we could further reduce the data  to include only results
closer to criticality. This moved a little bit the estimates, however,
did not diminish the systematic difference between the $t$ and $t'$
fits. We therefore conclude that in order to cure that problem most
likely  correction terms should be added in the
ansatz~eq.~(\ref{Eansatz}).  To this end, one would probably need more
or more precise data.

The slight mismatch of the result for $E_{\rm ns}$ obtained in  this
section with the result obtained from finite size scaling at the
critical point should  be  attributed to such corrections to scaling and
not to a failure of the theoretical prediction. 

Using the results for $E_{\rm ns}$, $C_{\rm ns}$ and $\nu$ obtained at 
the critical point one can compute the scaling amplitude from a single
energy value, just by solving eq.~(\ref{Eansatz}) with respect to
$A_{+}$ or $A_{-}$. The results are given in table~\ref{ampl}.

\begin{table}
\begin{center}
\begin{tabular}{|c|c|}
\hline 
$\beta$ & $A_{+}$  \\
\hline
0.2204  & 9.86(41)(18)\\
0.2205  & 9.85(40)(19)\\
0.2206  & 9.85(40)(19)\\
0.2207  & 9.84(39)(19)\\
0.2208  & 9.82(39)(20)\\
0.2209  & 9.82(39)(20)\\
0.2210  & 9.82(38)(20)\\
\hline
$\beta$ & $A_{-}$  \\
\hline
 0.2220 & 17.55(36)(46)\\
 0.2221 & 17.56(37)(39)\\
 0.2222 & 17.52(37)(37)\\
 0.2224 & 17.50(39)(36)\\
 0.2226 & 17.53(40)(35)\\
 0.2228 & 17.49(41)(34)\\
\hline
\end{tabular}
\parbox[t]{.85\textwidth}
{
\caption[ampl]
{\label{ampl}
Estimates for the amplitudes $A_{\pm}$ based on the estimates 
$E_{\rm ns}=0.330209(14)$,
$C_{\rm ns} = 11.1(8)$ and $\nu=0.6308(10)$.
The estimates are obtained by solving eq.~(\ref{Eansatz})
with respect to $A_{+}$ or $A_{-}$ a fixed $\beta$, 
assuming hyperscaling $\alpha = 2 - d \nu$.
The dominant sources of 
error in the resulting amplitudes are the errors of $C_{\rm ns}$
and $\nu$. These errors are displayed in the first and second 
brackets respectively.
}
}
\end{center}
\end{table}
The main sources of error in the amplitudes $A_{+}$ and $A_{-}$ computed
this way are  induced by the errors of $C_{\rm ns}$ and $\nu$. However,
when taking the  ratio $A_{+}/A_{-}$ from amplitudes computed at about
the same distance from $\beta_c$ the dependence on $\nu$ completely
cancels, and also the error by $C_{\rm ns}$ partially cancels. When
taking the amplitudes  obtained from the $\beta$-values closest to
$\beta_c$ we obtain  $A_{+}/A_{-} = 0.560(10)$, which is nicely
consistent with the result  that was obtained using only data with
$\beta\ne\beta_c$. 

Let us now make a comparison with a few results of the literature.
Estimates from  $\epsilon$-expansion, field theoretic calculations  in
$D=3$, high temperature expansions and from experiments are given
in~\cite{GuidaJustin}.  For the readers convenience, we reproduce part
of  that table in our table~\ref{amplL} and complete it with our 
present estimates.   Apparently, our estimates are larger than the
other ones cited in the table. However our most accurate estimate 
MC, (c) is consistent 
within error-bars with the most recent result form $\epsilon$-expansion 
\cite{GuidaJustin} and the results from renormalized perturbation theory
in three dimensions \cite{bagnuls,GuidaJustin}. We think that the 
disagreement with the estimate based on high and low temperature expansion
is most likely due to an under-estimation of the error in
ref. \cite{LiuFisher}.

\begin{table}
\begin{center}
\begin{tabular}{|l|l|l|l|}
\hline 
method                & $A_{+}/A_{-}$  & reference & year   \\
\hline 
$\epsilon$-expansion  & 0.524(10)      & \cite{NB}          & 85/86  \\ 
$\epsilon$-expansion  & 0.547(21)      & \cite{GuidaJustin} & 96     \\ 
field theory $D=3$    & 0.541(14)      & \cite{bagnuls}     & 87     \\ 
field theory $D=3$    & 0.536(19)      & \cite{GuidaJustin} & 96     \\
HT,LT series          & 0.523(9)       & \cite{LiuFisher}   & 89     \\ 
MC, (a)               & 0.550(12)      & this work          & 97     \\
MC, (b)               & 0.567(16)      & this work          & 97     \\ 
MC, (c)               & 0.560(10)      & this work          & 97     \\
\hline 
\end{tabular}
\parbox[t]{.85\textwidth}
{
\caption[amplL]
{\label{amplL}
Amplitude ratio estimates taken from the literatur and from this work.
The estimate (a) and (b) are the fit results quoted in table 
\ref{fitres}, and discussed in section~\ref{xxx}. 
Estimate (a) was obtained by including also information
from the data at the critical point, cf.\ the discussion 
at the end of section~\ref{xxx}.
Some estimates from experiments are
0.56(2) (binary mixtures), 0.49-0.53 (liquid-vapour systems), and
0.49-0.54 (magnetic systems), see ref.~\cite{privman_amp}.
}
}
\end{center}
\end{table}

\section{The universal constant $f_{\rm s} \xi^d$}

In this section we try to extract the numerical value of 
$f_{\rm s} \xi^d$
in both phases of the model. 
The values for the second moment correlation length  $\xi_{\rm 2nd} $
are taken from ref.~\cite{michelemartin}. 

The estimates for $f_{\rm s}$ at given $\beta-$values were obtained in the 
following way: We took $E-E_{\rm ns}-C_{\rm ns} (\beta-\beta_c) $
as an approximation
of the singular part of the energy.
The constants $E_{\rm ns}$ and $C_{\rm ns}$ were
taken from the combined energy and specific heat fit at the critical
point. Then we computed $f_{\rm s}$ as the integral over $\beta$ of the 
singular part of the energy.  We interpolated the singular part of energy 
for $\beta$-values not simulated with the scaling ansatz, where we use
$\nu=0.6308$ and the amplitude was computed from the closest $\beta$-value
simulated. The results are given in table~\ref{bloblo}.

In both phases we just have results for two $\beta$-values. Since the 
results of these two $\beta$-values nicely agree we regard the results 
$f_{\rm s} \xi^3 = 0.0355(15)$ and $f_{\rm s} \xi^3 = 0.0085(2)$  for the
high and  low temperature phase as reliable estimates for the critical
limits.

\begin{table}
\begin{center}
\begin{tabular}{|c|l|l|l|}
\hline 
$\beta$ & $\xi_{\rm 2nd}$ & $f_{s}$ & $ f_{s} \; \xi_{\rm 2nd}^3 $ \\
\hline
0.21931 & 8.760(5)&0.0000524(22) & 0.0352(15) \\
0.22020 &11.877(7)&0.0000212(9)  & 0.0355(15) \\
\hline
\hline
0.22311& 6.093(9)   & 0.0000377(9) &0.0085(2) \\
0.2240 & 4.509(6)   & 0.0000927(22)&0.0085(2)  \\
\hline
\end{tabular}
\parbox[t]{.85\textwidth}
{
\caption[bloblo]
{\label{bloblo}
In the second column we give results of ref.~\cite{michelemartin} for
the second moment correlation length $\xi_{\rm 2nd}$.  In the third
column we  give our estimate for the singular part of free energy
density $f_{\rm s}$,  while the  fourth column gives the resulting
estimate for the universal combination $f_{\rm s} \xi_{\rm 2nd}^3$.
}
}
\end{center}
\end{table}

Note that the result depends on the normalisation chosen here,
in particular we have chosen to take the free energy per link rather 
than per site.

\section{Conclusion}
By a  careful scaling and finite size scaling analysis of  energy and
specific heat data we obtained  estimates for various critical
quantities.  Taking into account the simplicity of the approach, 
especially the results for the exponents $\nu$ and  $\alpha$ are
remarkably precise.

\end{document}